# NFV and SDN - based Distributed IoT Gateway for Large-Scale Disaster Management

Carla Mouradian, Narjes Tahghigh Jahromi, Roch H. Glitho

*Abstract*— Large-scale disaster management applications are among the several realistic applications of the IoT. Fire detection and earthquake early warning applications are just two examples. Several IoT devices are used in such applications e.g., sensors and robots. These sensors and robots are usually heterogeneous. Moreover, in disaster scenarios, the existing communication infrastructure may become completely or partially destroyed, leaving mobile ad-hoc networks the only alternative to provide connectivity. Utilizing these applications raises new challenges such as the need for dynamic, flexible, and distributed gateways which can accommodate new applications and new IoT devices. Network Functions Virtualization (NFV) and Software Defined Networking (SDN) are emerging paradigms that can help to overcome these challenges. This paper leverages NFV and SDN to propose an architecture for on-the-fly distributed gateway provisioning in large-scale disaster management. In the proposed architecture, the gateway functions are provisioned as Virtual Network Functions (VNFs) that are chained on-the-fly in the IoT domain using SDN. A prototype is built and the performance results are presented.

## I. INTRODUCTION

Many large-scale disaster management applications rely on the Internet of Things (IoT), for example in fire detection and fighting, earthquake early warning and recovery, and flood monitoring. The IoT interconnects various addressable devices over the Internet in order to communicate and cooperate with each other when implementing specific applications [1]. Radio-Frequency IDentification (RFID) tags, sensors, and robots are examples of IoT devices. The IoT cannot stop disasters from happening but it can be very useful for disaster preparedness (e.g., prediction) and disaster recovery.

Different types of IoT devices are used in such large-scale disasters. For instance, sensors may be distributed throughout a forest to monitor the environmental conditions or measure earth movements before and during earthquakes, and robots can be used in search and rescue missions. Different types of sensors can be used, such as temperature, humidity, microwave, or infrared sensors. Similarly, robots with a wide range of capabilities can be used; some can penetrate rubble piles and find people beneath them, while others may be equipped with infrared cameras that transmit images back to the application. In addition, these IoT devices are usually heterogeneous, each with its own communication protocol and/or data formats.

To address the heterogeneity of IoT devices and their applications, gateways are needed to bridge the traditional communication networks and the IoT devices domain. Gateways are generally centralized and thus not practically feasible in the Mobile Ad-hoc Network (MANET) settings of large-scale disasters. Provisioning IoT gateways in large-scale disaster scenarios poses many challenges. For instance, traditional gateways lack dynamicity and flexibility. In addition, it is difficult and expensive to upgrade or reuse them. Network Functions Virtualization (NFV) [2] and Software Defined Networking (SDN) [3] can assist in overcoming these challenges.

NFV aims at virtualizing network services by decoupling the network functions from the underlying hardware. This could enable the on-the-fly, scalable, and elastic provisioning of the gateway by decomposing it into fine-grained modules (e.g., protocol converter, information model converter, metadata adaptor) and then implementing them as Virtual Network Functions (VNFs). SDN aims at splitting the control plane and the data plane in network elements in order to provide a flexible management of the forwarding behavior of those network elements. It can enable the easy on-the-fly chaining of these VNFs.

The major contribution of this paper is as follows: a distributed architecture for an IoT gateway based on NFV and SDN is proposed. The proposed architecture considers collocating the gateway functions with the IoT devices and reusing already deployed gateways. It also considers handling the traffic and chaining between the gateway functions dynamically. A high-level description of the proposed architecture that is composed of two planes is provided, and detailed description of each plane with its corresponding interfaces and procedures is presented. The proposed architecture is implemented as a proof of concepts in order to evaluate its viability and performance level.

The rest of the paper is organized as follows: Section II introduces a use case and discusses the requirements and related

C. Mouradian and N. Tahghigh Jahromi are with Concordia University, Montreal, QC H3G 1M8, Canada (e-mail: {ca_moura, n_tahghi}@encs.concordia.ca).

R. Glitho is with Concordia University, Montreal, QC H3G 1M8, Canada, and also with the Computer Science Programme, University of Western Cape, Cape Town 7535, South Africa (e-mail: glitho@ece.concordia.ca).

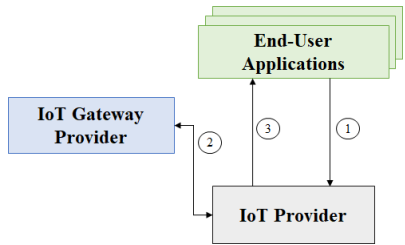

Fig. 1. The Proposed business model

work. Section III is devoted to the proposed architecture, including its modules, interfaces, and procedures. Section IV presents the implementation and investigates the performance evaluation, and Section V concludes the paper.

## II. USE CASE, REQUIREMENTS, & RELATED WORKS

### A. Use Case

An earthquake scenario can illustrate the motivation behind our work. One example is the large-scale earthquake that hit Kobe, Japan, in January 1996, measuring 6.9 magnitudes with more than 5000 dead and 13,000 injured. In such situations, early detection and fast recovery operations are required. *An earthquake early warning and recovery application* is considered that consists of two phases: an early warning phase and a recovery phase. In the first phase, the application monitors the environmental conditions by collecting environmental data such as the acceleration of physical objects, wind speed, etc. In the second phase, the application monitors the affected areas and deploys a fleet of heterogeneous search and rescue robots to assist in recovery procedures.

This application may have its own constraints in terms of communication protocols, data formats, etc. Moreover, the IoT devices may use different communication protocols and data formats. A gateway is needed to address this heterogeneity and allow interactions between the application and IoT devices.

Moreover, after an earthquake, there is an increasing possibility of fire. For instance, the earthquake in San Francisco in 1906 was followed by devastating fires that lasted for several days. In such cases, a *fire detection and fighting application* might consider adding new types of sensors with different capabilities, communication protocols, data formats, etc. to the infrastructure. These sensors allow the application to collect temperature, humidity, and $CO_2$ data in order to evaluate the contour and the intensity of any fire it detects and to dispatch firefighter robots accordingly.

### B. Requirements

First, the gateway should be deployed in a distributed manner due to the ad-hoc mode of disaster scenarios where there is no centralized or fixed infrastructure. Second, the gateway should be upgraded in a dynamic manner when new IoT devices are deployed. For instance, in case of fire, new sensors should be added to the infrastructure to send data to the fire detection and fighting application. This leads to the need to upgrade the gateway such that it can serve the newly added IoT devices. In addition, several applications should be able to use the same gateway. This may be required when, for instance, the sensors used by the fire detection and fighting application use the same protocol as the sensors used by the earthquake early warning and recovery application.

The gateway should support standard northbound and proprietary southbound interfaces. One example of a standard interface for sensors could be the widely used Sensor Markup Language (SenML) [4] carried over HTTP. SenML is designed to encode sensor measurements and device parameters and its extension could be used as a standard interface for robots. For instance, in [5], SenML is extended as a unified robots description model while in [6] SenML capabilities is extended to provide a uniform representation of sensors and robots and to control robots.

The architecture should also ensure that the execution of the gateway functions achieves performance levels similar to when they are executed in traditional gateways. Finally, the gateway should provide at least some key gateway functions, such as protocol conversion, information model conversion, data aggregation, and metadata adaptation.

### C. Related Works

Several works have proposed IoT gateway architecture. Some designed their architecture without considering the use of NFV/SDN technology. For instance, Datta *et al.* [7] propose a smart M2M gateway architecture to manage the huge volume of M2M devices and endpoints. They extended the capabilities of CoRE Link to add additional resource types for SenML units. However, such gateways cannot be upgraded on-the-fly when introducing new types of IoT devices or new applications.

Other works have investigated using NFV and/or SDN technology when designing IoT gateways. For instance, Li *et al.* [8] propose an IoT architecture based on SDN. In their proposed architecture, the gateways are SDN-enabled. They propose an architecture that allows introducing new applications through open programmable interfaces. It supports standard northbound interfaces using JSON and provides protocol conversion functionality as one of the gateway's key functionalities. However, the proposed gateway cannot be deployed over a MANET, as it does not have a distributed nature. In addition, it does not enable the same gateway to be used by more than one application.

TABLE I. SUMMARY OF THE RELATED WORKS

| Papers | Requirements | | | | | |
|---|---|---|---|---|---|---|
| | Distributed | Dynamic Upgrade | Reusability | Standard NB Interface | Good Performance | Key Gateway Functions |
| Datta *et al.* [7] | x | x | x | ✓ | N/A | ✓ |
| Li *et al.* [8] | x | ✓ | x | ✓ | x | ✓ |
| Ojo *et al.* [9] | x | ✓ | x | x | x | ✓ |
| Salman *et al.* [10] | x | ✓ | x | ✓ | x | ✓ |
| Mouradian *et al.* [11] | x | x | x | ✓ | ✓ | ✓ |

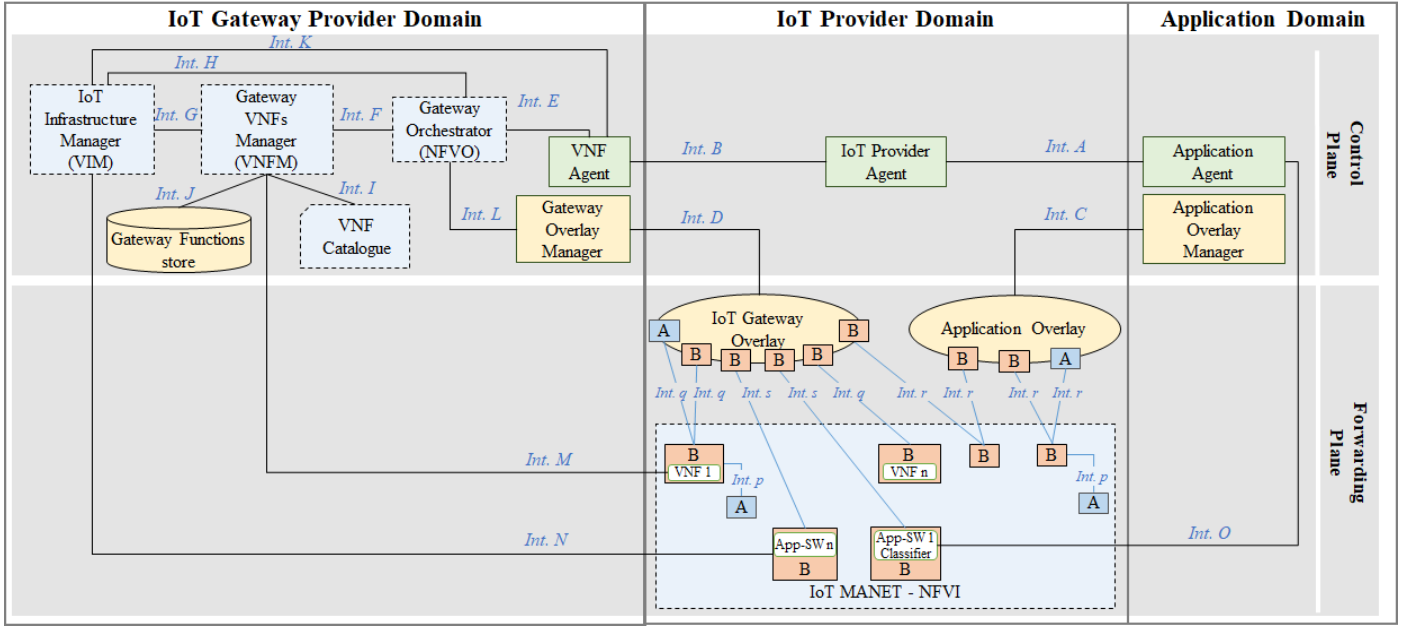

Fig. 2. The proposed distributed IoT gateway architecture

Ojo *et al.* [9] propose an SDN-IoT architecture coupled with NFV. Their goal is to address the scalability and the mobility issues in IoT networks. They replace traditional gateways with SDN gateways and implement the functionalities of the gateways as VNFs. The VNFs are SDN-enabled. This work supports heterogeneous IoT devices and provides key gateway functions. Also, the programmability feature of SDN allows the gateway to be updated dynamically. However, a drawback of this work is that the proposed architecture cannot be deployed over ad-hoc networks since it does not have a distributed architecture.

Salman *et al.* [10] propose a global IoT architecture leveraged with SDN. The proposed architecture inherits management and programmability capabilities from SDN and mobility capabilities from the fog. The gateways in the proposed architecture are SDF gateways that ensure interoperability between different communication protocols and heterogeneous networks, thereby providing key gateway functions. Their architecture also supports a standard northbound interface i.e., REST, and its programmability feature allows for the dynamic updating of the gateway. However, the SDF-gateway does not have a distributed architecture, and so it cannot be deployed over an IoT MANET.

Mouradian *et al.* [11] propose an NFV-based architecture for provisioning gateways for a Virtualized Wireless Sensor and Actuator Network (VWSAN). The VNFs implement protocol converter and information model processor to address the heterogeneity of IoT devices and the applications using them, thereby providing key gateway functionalities. The proposed architecture supports standard northbound interfaces and proprietary southbound interfaces; however, it deploys the entire gateway to a centralized location in the VWSAN domain, and thus it is not distributed. Moreover, it is not possible to upgrade the gateway dynamically when a new type of IoT device and/or application is deployed.

Table I provides a summary of the reviewed related works. For each reviewed paper, it shows the requirements which are met and the ones which are not met. We can notice that none of the reviewed papers propose a distributed architecture for the gateway. In addition, none of them proposes mechanisms for the reuse of the same existing gateway, or some of its functionalities by newly deployed gateways that serve new applications. Because of the dynamic nature of the MANETs, the proposed centralized IoT gateways become non-operational in the disaster area. In addition, dispatching brand new gateways when new applications are deployed requires expensive and complex operations. Hence, such gateways are not feasible in the MANET settings of large-scale disasters. In contrast, our proposed gateway is distributed and can be reused by more than one application.

III. ARCHITECTURE FOR A DISTRIBUTED IOT GATEWAY

This paper reuses the business model proposed in our previous work [11]. The specific actors and their relations are schematized in Fig. 1. The End-User Applications are the applications that require the services of an IoT Provider (Fig. 1, action 1). The IoT Gateway Provider provides on-the-fly VNFs, representing the gateway functions, to handle the heterogeneity of the applications and the IoT devices (action 2). The IoT Provider provides the IoT devices (i.e., sensors and the robots) required by the application to realize its service. A business agreement between the IoT Gateway Provider and the IoT Provider is assumed that allows the IoT Gateway Provider to manage the infrastructure on which the VNFs are deployed and executed.

It should be noted that in this paper it is assumed that there is a communication infrastructure that enables communication between the disaster site and disaster-safe areas, composed of IoT MANETs on the disaster site, and satellite mobile networks and cellular mobile networks to communicate with disaster-safe areas, as in reference [12]. Accordingly, the IoT provider can have a fixed node that can be carried on the relief vehicles during a MANET setup. The fixed node discovers and connects with the working cellular Base Station (BS). In fact, in practical situations, the cellular BS may be down at the central area on the disaster site. With luck, some BSs at a few distances away may be still working, allowing remote communication and information transfer between the IoT Provider in the disaster site area and the IoT Gateway Provider in the disaster-safe areas.

Some brief background information on NFV and SDN is given next. The architectural principles are then introduced, followed by a high-level description of the proposed distributed IoT gateway architecture. A detailed description of the control and forwarding planes, including the related architectural modules, interfaces, and procedures is discussed next. This section ends with the presentation of illustrative sequence diagrams.

*A. Background Information*

This subsection presents the background information that is relevant to the proposed architecture. The information covers the European Telecommunications Standards Institute (ETSI) reference architectural framework for NFV and the basics of SDN.

- The ETSI NFV [13] framework is made up of a set of three main components: VNFs, NFV Infrastructure (NFVI), and an NFV Management and Orchestration (MANO) framework. VNFs are the software implementation of given network functions. The NFVI provides hardware and software resources, including the computation, storage, and networking needed to deploy, manage, and execute VNFs. The NFV MANO framework enables the automated management of the VNFs by managing the NFVI and orchestrating the allocation of resources needed by the VNFs. It consists of three functional blocks: NFV Orchestrator (NFVO), VNF Manager (VNFM) and Virtualized Infrastructure Manager (VIM). The NFVO is responsible for the orchestration of the NFVI resources and the lifecycle management of the network services. The VNFM manages the lifecycle of the VNFs. The VIM is responsible for managing and controlling the NFVI.

- SDN aims at dissociating the control from the network forwarding elements, such as switches and routers, and centralizing the network intelligence at the controller. The architecture of SDN contains three planes: a management plane, a control plane, and a forwarding plane. The SDN application resides at the management plane. Its role is to communicate its requirements (e.g., the desired network behavior) to the control plane. This is done by defining a set of application policies and injecting them into the control plane. The control plane contains the SDN controller, whose responsibility is to translate the requirements of the SDN application to the forwarding plane. To that end, the SDN controller programs the forwarding plane by populating the SDN switches with well-defined flow entries (forwarding rules). The forwarding plane consists of forwarding elements such as switches and routers that allow traffic forwarding based on the flow entries that reflect the application's policies. NFV is highly complementary to SDN. Both are closely related technologies and are mutually beneficial (but not dependent). By using SDN, the network routers and switches can be dynamically programmed to steer the traffic through a set of VNFs.

*B. Architectural Principles*

The first architectural principle is that the application and the gateway are built as a P2P overlay, due to the MANET setting of the disaster scenario. The second principle is that two types of IoT devices are considered in the infrastructure, constrained (type A) and capable (type B). Constrained devices delegate some operations to the capable devices and capable devices can act on behalf of the constrained devices, as assumed in other works [14]. The third principle is that the IoT gateway functions are implemented as VNFs, and SDN is used to dynamically provision the paths between them once they are deployed.

*C. High-level Description of the Architecture*

The overall view of the proposed architecture is depicted in Fig. 2. It includes three domains: the Application Domain, the IoT Provider Domain, and the IoT Gateway Provider Domain. Each functionality of the IoT gateway is implemented as a VNF. It is important to note that the architecture, specifically the portion in the IoT Gateway Provider Domain, is aligned with ETSI NFV MANO framework [15]. In fact, it extends the MANO framework by adding new architectural modules.

The proposed architecture is layered over two planes: control and forwarding. The control plane handles the signaling procedure between the different domains, including the chaining of the gateway's VNFs. When an application requires an IoT service from the IoT Provider Domain (i.e., to receive sensor measurements and then deploy robots accordingly), a signaling procedure is conducted first to negotiate the service and exchange the necessary information. Different domains (i.e., the Application Domain, the IoT Provider Domain, and the IoT Gateway Provider Domain) are engaged in this service negotiation to obtain and deploy the appropriate gateway. The control plane also interacts with the forwarding plane by programming the application-level SDN switches to deliver the requested gateway to the application. Unlike traditional SDN switches, which are IP-level, the SDN Switches in the proposed architecture are application-level. This means the classification of a received packet should be done based on the header values of the application layer, similar to how it is done in [16].

The forwarding plane allows the flow of data through the path according to the control plane logic. It should be noted that all the interfaces of the proposed architecture are designed according to the Representation State Transfer (REST) principle.

TABLE II. PROSPECTIVE CHAINS BASED ON APPLICATION-LEVEL REQUIREMENTS

| End-user App. Requirements | | | IoT Devices Properties | | VNF Chains | Chain ID |
|---|---|---|---|---|---|---|
| Protocol | Info Model | Data Aggregation | Protocol | Info Model | | |
| HTTP | SenML | Average Data | CoAP | Raw Data | DA1, IMC1, PC1 | A |
| HTTP | SensorML | Average Data | HTTP | Raw Data | DA1, IMC2 | B |

TABLE III. APPLICATION-LEVEL FLOW TABLES

| App-level Switch | Match Field | Action |
|---|---|---|
| SW1 | Application: HTTP, SenML, Average data && IoT: Raw, CoAP | Insert Chain Id A Forward to SW2 |
| SW2 | Chain Id = A | Forward to SW3 |
| SW3 | Chain Id = A | Forward to SW4 |
| SW4 | Chain Id = A | Forward to "sensor a" |

*D. Detailed Description of the Control Plane*

Here the architectural modules involved in the signaling procedure along with their interfaces are presented, and some of the main procedures are described. It should be noted that some of the modules are reused from the ETSI NFV MANO framework, including the *NFVO* (i.e., *Gateway Orchestrator*), *VNFM* (i.e., *Gateway VNFs Manager*), *VIM* (i.e., *IoT Infrastructure Manager*), and the *VNF Catalogue*. Those modules are depicted with dashed lines in Fig. 2. The remaining modules are newly-introduced ones.

*1) Architectural Modules*

  *a) Modules in the Application Domain*

The *Application Agent* is in charge of the signaling procedure. It negotiates the use of IoT infrastructure. The *Application Overlay Manager* is responsible for the *Application Overlay* configuration, activation, and execution.

  *b) Modules in the IoT Provider Domain*

The *IoT Provider Agent* is responsible for requesting the desired gateway from the IoT Gateway Provider Domain.

  *c) Modules in the IoT Gateway Provider Domain*

The *VNF Agent* is responsible for analyzing the requested gateway's features, decomposing the request into a set of gateway VNFs that represent the gateway's functionalities, and requesting the execution of required orchestration plans. The SDN application resides in this module. It defines a set of chains which specify how the gateway VNFs are composed to fulfill the application's need(s).

As stated above, the proposed architecture reuses the *MANO* framework, including *NFVO*, *VNFM*, and *VIM*. The *NFVO* functionality is provided by the *Gateway Orchestrator*, which is in charge of orchestrating the *NFVI* resources and managing the lifecycle of the network services (i.e., the composition of VNFs). The *VNFM* functionality is provided by the *Gateway VNFs Manager*. This manager is responsible for the gateway's VNFs lifecycle management, including instantiation, maintenance, and termination. The *VIM* functionality is provided by the *IoT Infrastructure Manager*, which is responsible for resource allocations for the deployment and execution of the VNFs. The *SDN Controller* is co-located within the VIM. This is in accordance with ETSI SDN Usage in an NFV Architectural Framework [17]. According to this reference, one architectural option for the possible location of *SDN Controller* is to co-locate it with the VIM. This option is adopted in the proposed architecture. The *SDN Controller* establishes the path between the *VNFs* through which the sensor data and the commands to the robots traverse.

The *SDN Controller* is logically centralized. However, we envision it as being physically distributed to back-up controllers as suggested by references [18] [19] in order to incorporate fault tolerance mechanisms such as the ones described in these references. When it comes to scalability, we envision it as running in a virtualized environment in order to enable vertical and horizontal scalability. The reader should note that several existing SDN controllers (e.g., Floodlight [20]) do run in virtualized environments. The *SDN Controller* is connected to all the *Application-level SDN Switches* in the Forwarding Plane and programs them using an extended OpenFlow interface. The details of this interface are presented in the next sub-section.

The *Gateway Functions Store* includes the list of *VNFs* (i.e., gateway functions) that the IoT Gateway Provider Domain can provide. It is similar to the Network Function Store proposed by T-NOVA [21] which contains the VNFs provided by several third-party developers, published as independent entities and accompanied with their metadata. The *VNF Catalogue* represents a repository of all the on-boarded gateway VNF packages which are updated during their lifecycle management operation. The *Gateway Overlay Manager* is responsible for the *IoT Gateway Overlay* configuration, activation, and execution.

*2) Interfaces*

The general principles used to design the interfaces are presented first. This is followed by the description of the individual interfaces.

  *a) General Design Principles for the Interfaces*

As the SDN switches presented in the paper are application-level switches we first present the extensions made to the SDN interfaces. The general principles for the design of the other interfaces are presented after that.

**SDN interfaces**

In our proposed architecture, *Int. N* is the southbound interface and *Int. K* the northbound interface. OpenFlow and ForCES are the two standards currently used at the southbound interface [22]. However, they cannot convey application level information. In this paper, we have extended the most widely used standard (i.e., OpenFlow) as previously suggested in the literature [16][23]. An example of an extension is the addition of application-level fields to the match fields of the flow entries. The match field "OFB_IPV4_SRC" for instance is extended to support "source "OFB_Application_Level_Address_SRC" (in addition to "OFB_IPV4_SRC"). Yet another example, is the extension of the action "output" to output a message to an application-level address.

Contrary to the southbound API, there is currently no standard for the northbound interface [22]. Most SDN

controllers (e.g., OpenDaylight, Floodlight, etc.) do offer their own REST-based northbound APIs. However, like the southbound interface, the existing northbound interfaces do not cater to application-level features. If we take Floodlight, for instance, the resource "static flow pusher" REST API cannot be used as it stands to install flow table entries via REST API. We have therefore made extensions. An example is the extension of the matching field "ipv4_src" to include "application_level_address_src". Yet another example is the extension of "ipv4_dst" to include "application_level_address_dst".

**The Other Interfaces**

The other interfaces have been designed from scratch using the well-known RESTful Web services principles [24]. They all expose CRUD (i.e., Create, Read, Update, and Delete) operations. Table IV summarizes the proposed REST interface (i.e., *Int. E*) for the interactions between the *Gateway Orchestrator* module and the *VNF Agent* module. It defines resources on the *Gateway Orchestrator* "Orchestration Plan". This "Orchestration Plan" resource exposes a subset of the uniform interface to the *VNF Agent*. When the latter sends a POST request to the *Gateway Orchestrator* to request the execution of a new orchestration plan, the *Gateway Orchestrator* creates a new "Orchestration Plan" resource and sends the resource Uniform Resource Identifier (URI) to the *VNF Agent*. This URI is used to modify or get the status of an existing orchestration plan and to remove the resources of a specific orchestration plan.

  *b) Individual Interfaces Description*

*Int. A* is used by the Application Domain to request IoT service from the IoT Provider Domain. *Int. B* allows the IoT Provider Domain to request the desired gateway from the IoT Gateway Provider Domain. *Int. C* and *Int. D* are used for *Application Overlay* creation and *IoT Gateway Overlay* creation, respectively. *Int. E* is used by the *VNF Agent* to request the *NFVO* to execute the orchestration plan (deploy required *VNFs*, chain them, and create the gateway overlay).

*Int. F*, *Int. G*, and *Int. H* are reused from the ETSI NFV MANO framework; they represent Or-Vnfm, Vi-Vnfm, and Or-Vi, respectively. *Int. F* enables the instantiation, maintenance, and termination of the gateway's VNFs. *Int. G* enables the *NFVI* resource allocation for the gateway VNFs. *Int. H* enables the *NFVO* to monitor the *NFVI* resources. *Int. I* is a reference point in an ETSI NFV framework. It allows the *VNFM* to verify if the requested VNF is already deployed. *Int. J* is used to fetch the required VNFs from the *Gateway Function Store*. *Int. K* is the northbound interface of the *SDN Controller*, named Application Control Interface in ETSI SDN Usage in NFV [17]. *Int. L* is used by the *Gateway Orchestrator* to instruct the *Gateway Overlay Manager* to create the *IoT Gateway Overlay*.

*Int. M* and *Int. N* are reused from the ETSI NFV MANO framework; in MANO terminology these are Ve-Vnfm-vnf and Nf-Vi, respectively [13]. *Int. M* is for the lifecycle management of the *VNFs*, and *Int. N* represents the southbound API of the *SDN Controller*, named the SDN Resource Control Interface in ETSI SDN Usage in NFV [17]. *Int. O*, located between the

TABLE IV. EXAMPLE OF THE API OPERATIONS EXPOSED BY THE GATEWAY ORCHESTRATOR TO THE VNF AGENT (I.E., INT. E)

| REST Resource | Operation | HTTP Action and Resource URI |
|---|---|---|
| Orchestration Plan | Execute an orchestration plan | POST: /OrchestrationPlan |
| | Retrieve a specific orchestration plan | GET: /OrchestrationPlan/{Id} |
| | Retrieve all orchestration plan | GET:/OrchestrationPlan/all |
| | Remove an orchestration plan | DELTE: / OrchestrationPlan/{Id} |
| | Update an orchestration plan | PUT: / OrchestrationPlan/{Id} |

*Application Agent* and the flow classifier, is used to redirect the application request to the flow classifier.

  *3) Procedures*

The proposed architecture, as part of the signaling procedure, includes the following two procedures: IoT gateway provisioning and application provisioning. IoT gateway provisioning includes the IoT gateway request and the IoT gateway orchestration. IoT gateway orchestration refers to the IoT gateway deployment, IoT gateway chaining, and IoT gateway overlay creation. For the application provisioning procedure, this paper focuses on the application overlay creation phase.

Next, the gateway orchestration procedure (Fig. 3-a) with its three phases: deployment, chaining, and overlay creation, is described.

  *a) IoT Gateway Deployment*

The process starts when the IoT Gateway Provider Domain receives a gateway request from the IoT Provider domain (through Int. B). The *VNF Agent* instructs the *Gateway Orchestrator* to execute the orchestration plan through *Int. E*. The *Gateway VNFs Manager* first checks if the *VNFs* needed for the requested gateway are already present in the IoT Provider Domain. If not, the *Gateway Orchestrator* discovers the capable IoT devices in the IoT Provider Domain (i.e., type B) along with their capabilities and features, such as energy level, response time, location, etc., and maintains a clear view of the network topology. The *Gateway VNFs Manager* then finds the requested *VNFs* in the *Gateway Functions Store* (through Int. J) and instantiates and dispatches them (through Int. M).

  *b) IoT Gateway Chaining*

Once the *VNFs* are deployed, the *SDN Controller* (co-located with the *IoT Infrastructure Manager)* programs the *Application-level SDN Switches* (via Int. N). The *SDN Controller* populates a set of application-level flow entries in the *Application-level Switches* based on the chains defined by the SDN Application in the *VNF Agent*. The SDN application injects these entries in the *SDN Controller* via Int. K. According to these entries, a path is established between the application and the IoT devices through which the data from IoT devices traverse to the application.

  *c) IoT Gateway Overlay Creation*

This process is initiated when the *Gateway Overlay Manager* receives a request from the *Gateway Orchestrator* to create the *IoT Gateway Overlay* (through Int. L). The *Gateway*

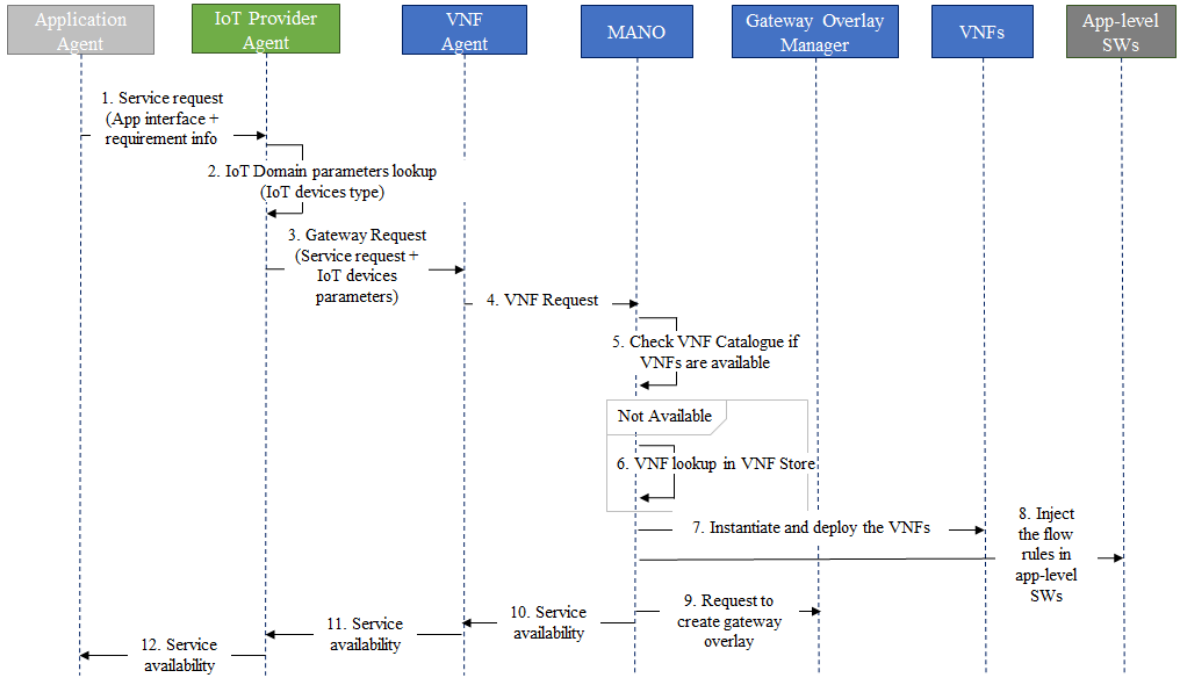

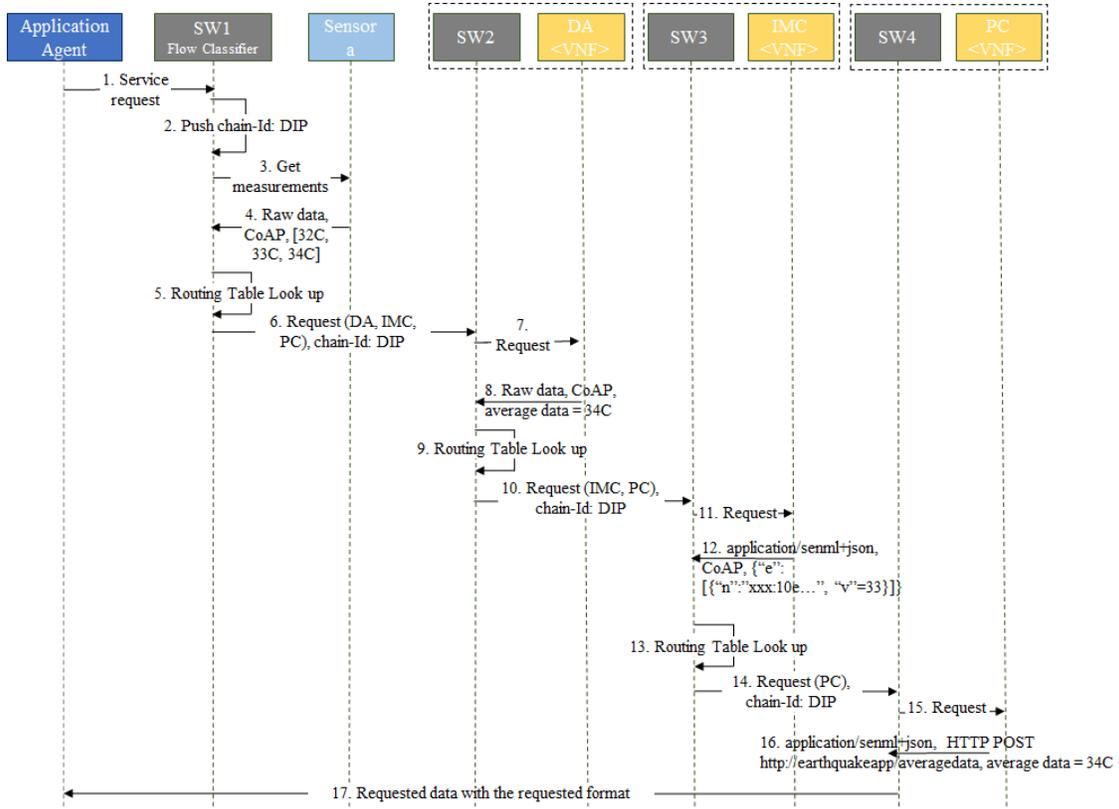

Fig. 3. Sequence Diagram for the Gateway Deployment and Chaining
(a) IoT Gateway Provisioning Procedure
(b) End-to-End Scenario of the Actual Flow of Data

*Overlay Manager* first configures the *IoT Gateway Overlay* between the selected type B devices. It then activates the overlay, where the selected IoT devices receive an overlay join request (Int. D).

The proposed architecture in Fig. 2 includes two overlays built on top of the IoT MANET: the *IoT Gateway Overlay* and the *Application Overlay*. They co-exist simultaneously; each may have its own overlay protocol for message exchange. They share nodes and underlying network links. In order to allow these overlays to interact and cooperate with each other, the proposed approach uses co-located nodes, (nodes that belong to the two overlays) to enable inter-overlay routing and reduce traffic [25]. Every message received by a co-located node can be forwarded to the other overlay the node belongs to. Using super-peers is another approach; however, it leads to costly merging mechanisms [26] and [27].

*E. Detailed Description of the Forwarding Plane*

The architectural modules and the interface of this plane are described below. The *NFVI* is reused from the ETSI NFV architectural framework [13] and is shown with dashed lines in Fig. 2.

*1) Architectural Modules*

All the modules described in this section are in the IoT Provider Domain. The *NFVI* is able to host the *VNFs* deployed over IoT devices. The *VNFs* are the software instances of the gateway functions.

Type A IoT devices rely on type B devices to join the overlay. They send data to the *IoT Gateway Overlay* through type B devices. Type B devices can execute one or more function of the gateway, and they can represent themselves and/or a type A device in the *IoT Gateway Overlay*. Similarly, type B devices can join the *Application Overlay* on behalf of type A devices. In addition, the same type B device may belong to both overlays.

The *Application-level SDN Switches* are programmable by the *SDN Controller*. The *Application-level SDN Switches* in the proposed architecture are placed on a computing hardware in the *NFVI* (as one of the options presented in [17]) in which some type B devices act as SDN switches. It is worth noting that some industrial projects are working on using application-layer intelligence in an SDN environment (i.e., a white paper [28]).

*2) Interfaces*

*Int. p* is used to exchange control data between type A and type B IoT devices. *Int. q* is used to send the sensor data to the *IoT Gateway Overlay*. *Int. r* is used to send the data received from the IoT devices executing an application's task to the *Application Overlay*. *Int. s* allows communication between the *Application-level SDN Switches* in order to establish the path between them. It also allows pushing/retrieving of the sensor data between the *Application-level SDN Switches* and the *VNFs*.

*F. Illustrative Sequence Diagram*

Fig. 3 illustrates a sequence diagram of the interactions of different architectural modules during the signaling procedure for gateway provisioning (gateway request and gateway orchestration). It also illustrates an end-to-end scenario of the actual flow of data when sending data from the sensors to the application. A fire detection and fighting application is considered, which wants to collect data from temperature sensors to detect prospective fires after an earthquake, and can then deploy firefighting robots in case of a disaster.

The gateway provisioning procedure is shown in Fig. 3-a. MANO in the figure represents the *Gateway Orchestrator* (i.e., NFVO), the *Gateway VNFs Manager* (i.e., VNFM), and the *IoT Infrastructure Manager* (i.e., VIM). During the deployment phase, the *VNFs* are selected in the IoT Gateway Provider Domain. This selection is done based on the application's interface description and requirements (Fig. 3-a action 1), e.g., SenML over HTTP, average data and IoT device specifications (action 2-3), e.g., Virtenio Sensors with CoAP protocol that send raw data. The IoT devices' specifications are found in a repository held by the IoT Provider Domain.

It is assumed that the requested Gateway's VNFs implement the following functions: (1) Data Aggregator (DA) to send data over a specific threshold; (2) Protocol Converter (PC) to convert the data received from IoT devices to the appropriate model supported by the application and vice versa; and (3) Information Model Converter (IMC) to convert a model from one to another. We do acknowledge the fact that converting a protocol X (or an information model X) into a protocol Y (or an information model Y) is not always feasible. Consequently, the IoT Gateway Provider provisions the related VNFs only when the conversion is feasible. It is important to note that when the required *VNFs* are not found in the *Gateway Functions Store*, a service unavailability notification is sent to the *Application Agent*, to either cancel the negotiation or resume signaling after a certain time.

The selected VNFs are instantiated and deployed in the IoT Provider Domain (action 8). The *IoT Infrastructure Manager* next injects the flow entries in the *Application-level SDN Switches* (action 9). Some examples of such entries are listed in Table III. The *IoT Gateway Overlay* is then created in the IoT Provider Domain (action 10). Finally, the *Application Agent* receives a notification about service availability through the *VNF Agent* and the *IoT Provider Agent* (actions 11-13) in which the *Application Agent* is instructed to contact the flow classifier to collect data from the sensors. Table II demonstrates two prospective chains that could be defined for these VNFs based on the end-user application preference and the IoT devices' properties.

The end-to-end scenario of the forwarding plane is shown in Fig. 3-b. The *Application Agent* sends its request to the SW1 (action 1) which performs the flow classification according to Table III. It then pushes the chain-Id on the request (action 2). This chain-Id indicates that DA, IMC, and PC are needed to deliver the required service to the application. SW1, according to the entries in its routing table, (i.e., Table III) collects measurements from sensors (e.g., sensor x, type A) and sends the request to the second switch (SW2) (action 3-6). The SW2 sends the request to the DA VNF to aggregate the data (actions 7-8). The same applies to SW3 and SW4 (actions 9-16). Finally, the requested data is sent back to the *Application Agent* (action 17) through the *Application Overlay*.

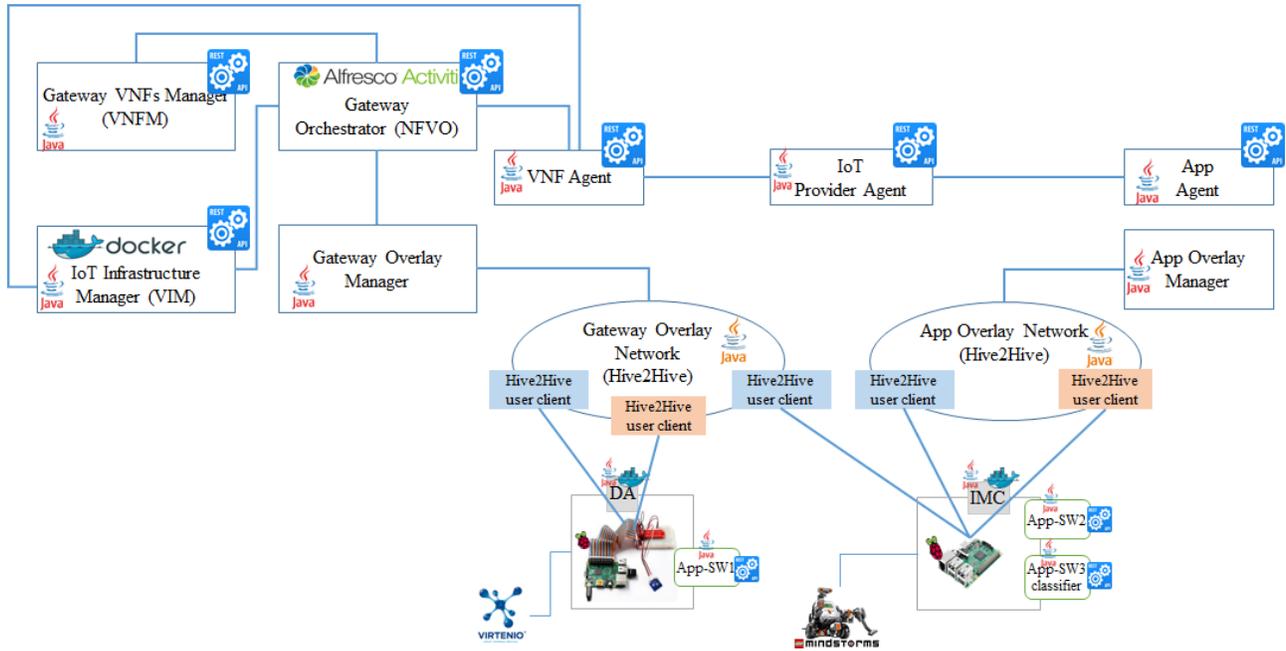

Fig. 4. Implementation architecture

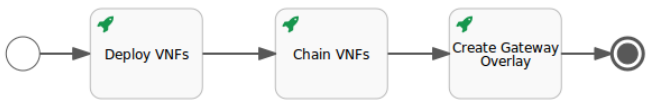

Fig. 5. IoT gateway orchestration plan

## IV. IMPLEMENTATION AND EXPERIMENTATIONS

### A. Implementation Scenario

For the prototype, the recovery phase of the scenario presented in Section II.A is implemented. This phase is as follows; an earthquake recovery application collects the data of sound sensors deployed in the affected areas. These sensors can detect voices or other sounds of possible human presence through the ruins and inform the earthquake recovery application. In order to communicate with these IoT devices, the application needs a gateway for handling the different types of communication interfaces. Sometime later, a fire detection and fighting application that needs to be notified when a fire occurs adds temperature sensors that can detect fires. It then deploys a fleet of robots that can detect extinguishers and grab one in order to suppress the fire. Accordingly, the gateway needs to be upgraded in a dynamic manner such that it can serve the newly added IoT devices.

Three different types of IoT devices are used. Sunfounder[1] sensors are deployed over the Raspberry Pi[2] (RPi) to detect sounds, Virtenio[3] sensors are deployed to detect fires, and Lego Mindstorms[4] robots with specialized arms are utilized to grab extinguishers and suppress fires. A ball is used to simulate the extinguisher.

The earthquake recovery application has one application component: a human presence detector. To that end, Sunfounder sensors on RPis send their raw measurements over HTTP. Protocol. This data is first processed by a DA (to send only the sounds of any possible human beings) and then processed by the IMC (mapped to SenML format).

The fire detection and fighting application has two application components: fire detection and robot deployment. Virtenio sensors send their raw measurements over CoAP protocol. These measurements are processed by a DA (to send data over a specific threshold) followed by an IMC (mapped to SenML format), and finally by a PC (encoded in HTTP protocol). Accordingly, the same IMC can be reused by both applications. Finally, in order for the application to send commands to the robots, the IMC and the PC convert the HTTP request received into LeJOS JAVA API command that implements the LCP.

In this prototype, it is assumed that an ad-hoc network is already built and that the connection between the involved nodes has been established. It is also assumed that a fixed node in the IoT provider is carried on the relief vehicles during MANET setup and is part of the gateway and the application overlay.

### B. Implementation Architecture

The validation prototype is implemented according to the architecture depicted in Fig. 4. In this implementation, container-based gateway orchestration is adopted. Containers

---

[1] sunfounder.com/modules/sensor-module.html
[2] raspberrypi.org/
[3] virtenio.com/
[4] lego.com/en-us/mindstorms

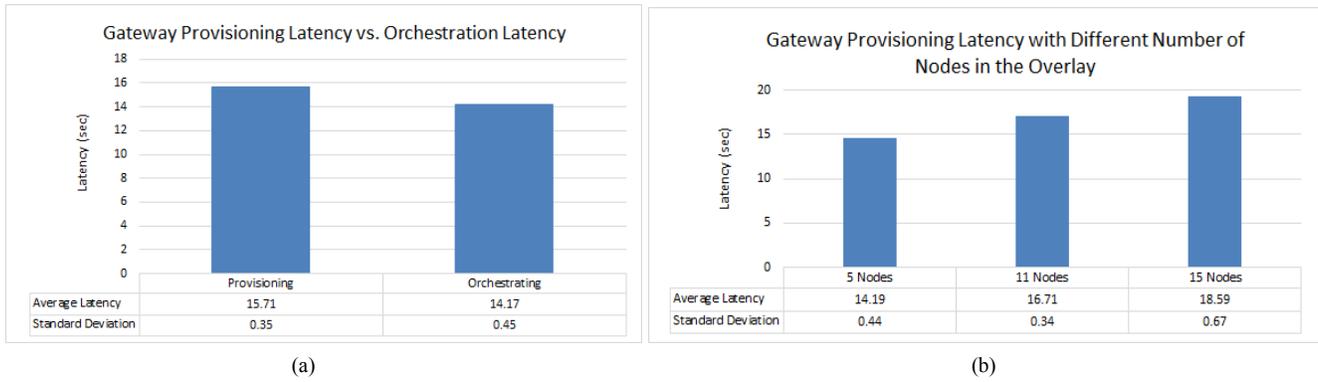

Fig. 6. Gateway provisioning latency

(a) Provisioning latency vs. orchestration latency
(b) Provisioning latency with different number of VNFs and Applictaion-level SDN Switches

are lightweight, stand-alone, and modular. In addition, they allow rapid configuration and deployment. Existing open source MANO solutions are not used in this prototype as they do not support container-based orchestrations.

For the *Gateway Orchestrator* (i.e., NFVO), Alfresco Process Services[5], an enterprise Business Process Management (BPM) solution is used. It allows the creation of process definitions and orchestration plans using the capabilities of Business Process Model Notation (BPMN). It also exposes REST API to external entities to execute the orchestration plan. Fig. 5 shows an example of the orchestration plan used for the new deployment of the gateway. In this plan, the required VNFs for the gateway are first deployed, then the VNFs are chained using SDN, and finally, the *IoT Gateway Overlay* is created.

The application-level switches are implemented using Java libraries. These switches expose a REST API implemented as Java Restlet framework to the *SDN Controller* through the fixed node in the IoT Provider domain for handling application-level flow entries. For the SDN Controller, since existing open source SDN controllers do not support our proposed extended features, we implemented it as a simple REST API using the Restlet framework.

The *IoT Gateway Overlay* is created by the *Gateway Overlay Manager*. Hive2Hive[6] API is used, which provides a free, open-source, distributed, and scalable solution for distributed P2P networks. It also provides for headless deployment (e.g., on an RPi) and is configurable and customizable. The overlay is created by the fixed node in the IoT Provider domain as a master client by advertising its address and creating a user profile. The remaining nodes join the overlay network as user clients by registering to the user profile.

The VNFs are implemented using Java libraries and packaged in Docker containers. They are pushed to the DockerHub repository. The container images are pulled from the repository through the fixed node in the IoT Provider domain. The *Application-level SDN Switches* and the VNFs communicate with each other in the overlay using the Hive2Hive framework.

Two VNFs (i.e., DA and IMC) and three Application-level SDN Switches are randomly placed on the two RPis. It should be noted that a VNF placement algorithm can be adapted to deploy the VNFs in the optimal location in the network (e.g., [29].

The remaining architectural modules are modeled as RESTful web services using a Java Restlet framework.

### C. Setup

The IoT Gateway Provider domain runs on a 64-bit laptop with Ubuntu 14.04.5 LTS. The Application domain runs on a laptop with Intel® Xeon® CPU clocked at 2.67 GHz and a 6GB RAM with 64-bit Windows 7 Enterprise. The fixed node of the IoT Provider domain runs on a 64-bit laptop with Ubuntu 14.04.5 LTS. The RPis run a Raspbian OS which is based on Debian OS. The first RPi hosts two Hive2Hive user clients (i.e., one on behalf of a Virtenio sensor and one on behalf of the DA VNF). The second RPi hosts three Hive2Hive user clients (i.e., one on behalf of a LEGO Mindstorms robot, one on behalf of an IMC VNF, and one representing the human presence detector component of the earthquake application.

### D. Performance Evaluations

#### 1) Performance Metrics

The performance metrics utilized to evaluate the performance of the proposed architecture are:

**Gateway provisioning latency** – Measured from the time the *Application Agent* sends a request to obtain sensor measurements to the time the gateway is deployed and chained in the IoT domain. Experiments with a different number of VNFs, SDN switches, and overlay nodes are conducted.

**Orchestration latency** – Measured from the time the orchestration request to the *Gateway Orchestrator* is initiated to the time the acknowledgment of orchestration is received. The orchestration process, as discussed in the previous section, includes the IoT gateway deployment, IoT gateway chaining, and IoT gateway overlay creation. In addition, the orchestration latency is measured when a request to upgrade an already-deployed gateway is received. To that end, a scenario where new sensors are added by the fire detection and fighting application is assumed. According to this scenario, a gateway with three VNFs is required (i.e., DA, IMC, PC), with one of the VNFs (i.e., IMC) already deployed in the IoT Provider Domain.

---

[5] alfresco.com

[6] hive2hive.com

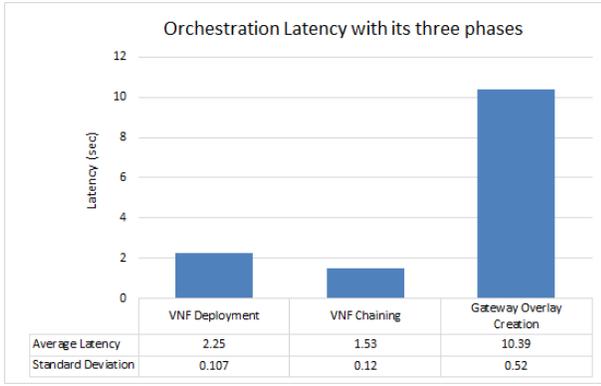
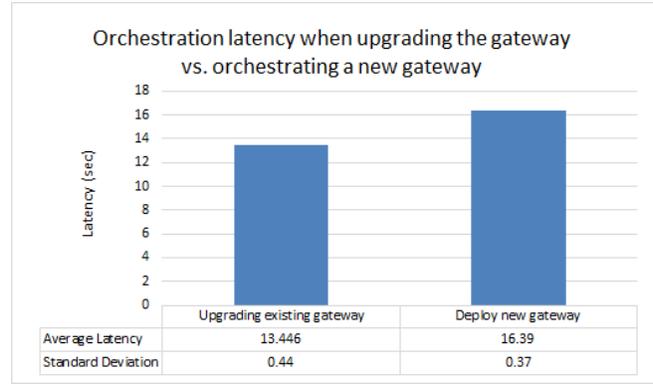

Fig. 7. Orchestration latency of the proposed gateway
(a) Orchestration latency of each phase of the orchestration plan
(b) Orchestration latency of upgrading the proposed gateway vs. orchestrating a new gateway

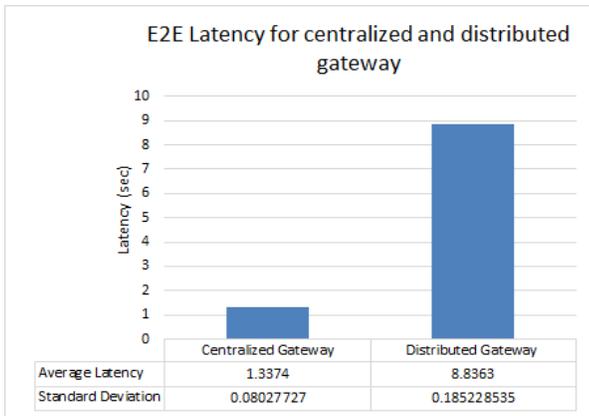
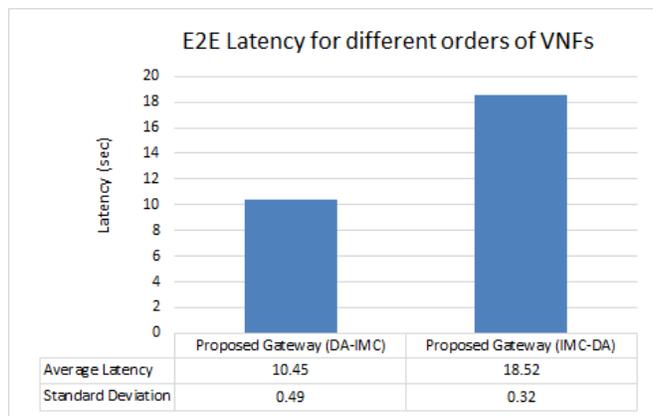

Fig. 8. E2E Latency of the proposed gateway
(a) E2E delay for centralized gateway vs proposed distributed gateway
(b) E2E delay of the proposed gateway for different orders of VNFs

**End to End (E2E) delay** - Measured from the time the IoT devices send their data to the time the requested data is obtained by the application. Here, the order of the VNFs is varied to show the effect of changing the order of the VNFs on the E2E delay. The E2E delays for both fixed/centralized gateways and the proposed distributed gateway are also calculated.

*2) Results and Discussions*

This section discusses the performance results obtained, beginning with the provisioning latency.

*Test Case 1: Provisioning latency*

Fig. 6-a depicts the IoT gateway provisioning latency and the orchestration latency, which is a sub-phase of the provisioning procedure. Both average latency and the standard deviation are provided for 10 consecutive experiments. The average provisioning latency is 15.71 sec. This shows the efficiency of using NFV and SDN-based gateways compared to traditional gateways. The latter imposes lengthy deployment time, requires additional configurations for physical interfaces to implement chains, etc. [30]. Also, it can be observed from the graph that the average orchestration latency (14.17 sec) takes up most of the time of the provisioning procedure.

In order to conduct an accurate insight into how the system behaves depending on the number of VNFs, SDN switches, and overlay nodes, several cases are carried out. A linear topology is used for the Application-level SDN Switches. We gradually increase the number of instances of each VNF while considering a load balancer VNF for each group of VNFs of the same type to equally distribute the load among them. Therefore, the case with 6 VNFs for instance, include 3 instances of each VNF (i.e., Data Aggregator and Information Model Converter), 2 load balancers, and 7 application-level SDN switches. This leads to a gateway overlay with 15 nodes. A Similar number of overlay nodes for disaster management scenarios are considered in the literature [31]. It can be observed in Fig. 6-b that by increasing the number of VNFs and the SDN switches the provisioning latency behaves almost at a slight constant rate, linearly. This is because when the number of nodes in the overlay is increased, the system experiences a very slight increase in the overlay creation latency. This is since all the nodes (i.e., user clients in Hive2Hive terminology) need to register to the user profile created and advertised by the fixed node (i.e., master client) and join the overlay.

Next, the various phases of the orchestration are measured to get a more detailed insight into the orchestration latency.

*Test Case 2: Orchestration latency*

Fig. 7-a shows the latency of the orchestration, including IoT gateway deployment, IoT gateway chaining, and IoT gateway overlay creation over 10 consecutive measurements. Both average latency and the standard deviation are provided. It can be noted that the overlay creation takes up most of the orchestration time, which indicates that overlay creation imposes an overhead on the overall management and orchestration procedures in terms of latency.

One reason for the time required for overlay creation is that with overlay networks, an additional intermediate level is added between the nodes in the physical network infrastructure. However, despite the overhead, the experimental results show that the overhead associated with the overlay network is not a significant factor compared with the considerable gain of the approach. Overlays play an important role in MANETs where nodes join and leave the group dynamically. The additional layer prevents interference with the existing protocols in the underlying heterogeneous environment of MANET.

Fig. 7-b shows the orchestration latency associated with the procedure of upgrading the gateway vs deploying a new gateway. The proposed architecture allows the upgrading of a pre-deployed gateway for which the orchestration latency is 15.06 sec. In contrast, the orchestration latency associated with the procedure of deploying a completely a new gateway (when upgrading the gateway is not supported) is 24.51 sec. It can be observed that the proposed architecture's feature of allowing the gateway's upgrading decreases the orchestration latency by 38.56%.

*Test Case 3: End to End delay*

In Fig. 8-a, the E2E delay of the proposed distributed gateway with two VNFs (DA-IMC) is compared to a fixed centralized gateway. The fixed gateway aggregates the received data and converts the information model into two tightly coupled functionalities. The average latency for the proposed distributed gateway (i.e., 8.8 sec) is higher than the average latency for a centralized fixed gateway (i.e., 1.3 sec). This is basically because in a centralized gateway the communication latency between different functionalities is eliminated.

However, in the implementation, a linear topology is adopted for the SDN switches; other topologies could be investigated to see if they can reduce this latency. Eliminating the SDN switches and making the VNFs SDN enabled could also reduce this latency, but this might make the VNFs more demanding in terms of processing and storage and may not fit resource-constrained environments. Using other P2P frameworks could also be investigated to see if they can reduce this latency.

In Fig. 8-b, the E2E delay of the proposed distributed gateway with two VNFs (DA-IMC) is measured for two valid chaining options of the VNFs. Five consecutive measurements are assumed to be sent from the sound sensor, where the last measurement indicates a possible human being sound. In the first chain (i.e., DA-IMC), the DA receives all the measurements and sends only possible human being sound which is converted to the appropriate model at the IMC. In the second case (i.e., IMC-DA), all the received measurements from the sensors are converted to the model supported by the application (i.e., SenML) and then filtered at the DA to send only the possible human being sound. It should be noted that both cases lead to the same relevant result at the application. The average latency in the first chain (i.e., DA-IMC) is 10.45 sec, while in the second chain (i.e., IMC-DA) it is 18.52 sec.

It is clear that the performance of the gateway in terms of latency is directly affected by the order in which the VNFs are composed and processed. This order has a significant impact on the amount of processing capacity required for each VNF. And, the amount of the required capacity depends on the amount of data handled by that VNF instance. Different chain composition algorithms can be integrated to improve the performance of the gateway e.g., [29].

## V. CONCLUSIONS

This paper proposes an architecture for on-the-fly distributed gateway provisioning in disaster management using NFV and SDN technologies. These technologies make it possible to address the challenges of traditional gateways, such as dynamicity and flexibility. NFV allows upgrading the pre-existing gateway and deploying the gateway functions anywhere anytime, and SDN enables reusing the same gateway functions in different flows for different applications. The gateway functionalities are provisioned as VNFs and are chained dynamically using the application-level SDN switches. The IoT gateway is built as a P2P overlay taking into consideration the MANET settings of the disaster management scenarios.

A prototype of the proposed architecture is provided and a set of experiments are conducted to evaluate the architecture. The results show that building the IoT gateway as a P2P overlay imposes an overhead on the overall management and orchestration procedures but it produces a considerable gain. The performances of distributed and centralized approaches are also analyzed and the effect of the order of the VNFs on the overall E2E delay is investigated. The results show the advantages of using on-the-fly provisioning of IoT gateway and the possibility of reusing and updating a pre-existing gateway.

There are several potential items for future work. One example is the design of a VNF placement algorithm. Such algorithms need to be adapted to the context of IoT, taking into consideration the mobility of the IoT devices hosting the VNFs while ensuring the QoS.

ACKNOWLEDGMENT

The work is partially supported by the Canadian Natural Science and Engineering Council (NSERC) through a Discovery Grant.